\documentclass[%
 reprint,
 amsmath,amssymb,
 aps,
]{revtex4-2}

\usepackage{graphicx}
\usepackage{dcolumn}
\usepackage{bm}

\usepackage[usenames]{color}
\usepackage[normalem]{ulem}

\begin{document}

\preprint{APS/123-QED}

\title{\textbf{Spintronic THz emitters based on NiCu alloys.} 
}%

\author{Evgeny A. Karashtin$^{1,2,3}$}
 \email{eugenk@ipmras.ru}
\author{Igor Yu. Pashen'kin$^{1}$}%
\author{Anastasiya V. Gorbatova$^{3}$}
\author{Ekaterina D. Lebedeva$^{3}$}
\author{Pavel Yu. Avdeev$^{3}$}
\author{Nikita V. Bezvikonnyi$^{3}$}
\author{Arseniy M. Buryakov$^{3}$}

\affiliation{$^{1}$Institute for Physics of Microstructures RAS, Nizhny Novgorod,  Russia}
\affiliation{$^{2}$Lobachevsky State University of Nizhny Novgorod, Nizhny Novgorod, Russia}
\affiliation{$^{3}$ MIREA – Russian Technological University, Moscow, Russia}%

\date{\today}

\begin{abstract}
We study THz emission from ferromagnet / nonmagnetic material (FM/NM) spintronic nanostructures in which the $Ni_xCu_{1-x}$ alloy with different $x$ is used as an FM, an NM, or both layers. The stoichiometric composition of the NiCu alloys standing at two positions (we denote it as [FM] or [PM]) is chosen so that it is ferromagnetic at room temperature in the case it is used as the FM layer, and is paramagnetic at room temperature for the NM layer. Besides, we choose the nickel ratio $x$ close to each other for both [FM] and [PM] types of the alloy (the difference is only $10\%$). We show that although NiCu[PM] does not contain heavy metal it acts as an effective converter of spin current into the electric one in our structure showing only 2.8 times smaller efficiency than Pt. Besides, the NiCu[FM] alloy, despite having quite small Curie temperature (approximately $65 ^\circ C$), acts as an effective spin source having the efficiency only 2 times smaller than Co in similar structures. This shows up the importance of boundary matching in the spintronic THz sources. Our NiCu-based THz sources reveal a possibility of effective thermally induced control of emission of THz radiation due to a unique combination of high emission rate and relatively small Curie temperature.  
\end{abstract}

\maketitle


\section{\label{Intro} Introduction}

Spintronic terahertz emitters based on ferromagnet / nonmagnetic layer (FM / NM) heterostructures are simple and powerful sources of THz radiation \cite{Kampfrath2013}. They combine the advantages of compact device dimensions, low-cost fabrication process, broad bandwidth and relatively high efficiency \cite{Bull2021, Seifert2016}. Although there are still debates on the importance of the ultrafast demagnetization process on the work of spintronic THz emitters, the conventional physical principle of their work is based on two main steps: The excitation of the spin current from the FM layer to the NM layer, and the conversion of this spin current to the electric current that irradiates electromagnetic wave (usually through the inverse spin Hall effect \cite{Averkiev, Bakun, Sinova, Valenzuela, Saitoh2006} or the Rashba-Edelstein effect \cite{Fert2013}). The effectiveness of these two processes has been the main point of the effort en route to optimize the emitters. The main steps on this way include the choice of a ferromagnetic material with high degree of spin polarization (usually Fe, Co, or CoFeB) as an FM and a heavy metal layer with high spin-orbit coupling (SOC; usually Pt, Pd, or W) as as the NM layer. The influence of the boundary between FM and NM on the injection of spin current is also discussed \cite{Mikhailovsky, Karashtin_Si}. Certain attempts to create a transition layer consisting of a FM-NM alloy are performed that show up a possibility to increase the THz emission rate by matching the layers' boundaries.

The intermetallic NiCu alloys reveal a unique combination of physical and magnetic properties. Although they do not contain heavy metal atoms, the alloys have relatively high spin Hall angle ($\theta_{SH} \approx 4.1\%$ \cite{Varotto}) which is comparable to that of Pt ($\theta_{SH} \approx 3-10\%$ \cite{Wang}). The Curie temperature of the $Ni_xCu_{1-x}$ alloy depends on $x$ and may be either greater than room temperature (in this case the alloy is ferromagnetic hence we refer to it as NiCu[FM]) or lower than it (the alloy is paramagnetic at room temperature, and we call it NiCu[PM]) \cite{Cheng}. Thus, it is possible to realize a structure in which two adjacent NiCu layers have different properties and act as both the FM and NM layers of the spintronic THz emitter. On the one hand, this leads to a good boundary matching. On the other hand, the ability to change the Curie temperature which is slightly greater than room temperature by varying $x$ in the NiCu[FM] layer opens up a possibility of thermoinduced control of its magnetization \cite{Korenivski}.

Current paper is devoted to a problem of use of NiCu with different alloy composition both as FM and NM layers of the spintronic THz emitter. We investigate terahertz emission from the described heterostructures $Ni_xCu_{1-x}$/$Ni_yCu_{1-y}$ (or NiCu[FM]/NiCu[PM]) in which $x = 0.78$ and $y = 0.68$. We compare our results to three reference structures: traditional Co/Pt, Co/NiCu[PM] in which spin-to-charge conversion is realized in a NiCu alloy, and NiCu[FM]/Pt in which NiCu is the source of spin and show that the boundary match is an important factor in the emission of THz radiation by the investigated structures. 

The paper is organized as follows. First, we describe sample preparation and setup of the experiment. Next, the THz emission rate is compared for different structures. After that, the results of temperature dependent measurements are described. We show how the THz signal can be manipulated by pump fluence and estimate the sample heating induced by it. A brief discussion of the obtained results showing the importance of boundary match is summarized in the Conclusion.

\section{\label{Setup} Samples and experimental setup}
We study four samples: a(6nm)/$Ni_{0.68}Cu_{0.32}$(6nm) is the main object of our study; Co(3nm)/Pt(3nm) which is one of the most effective THz emitters is the main reference sample; $Ni_{0.78}Cu_{0.22}$(6nm)/Pt(3nm) and Co(3nm)/$Ni_{0.68}Cu_{0.32}$(6nm) are investigated for comparison and in order to determine the role of boundary match in THz emission. The 3~nm Co and Pt layers' thickness is optimal for our Co/Pt emitter efficiency. The $Ni_{0.68}Cu_{0.32}$ (NiCu[PM]) layer thickness is chosen in order to increase its spin-to-charge conversion efficiency \cite{Cheng}. The $Ni_{0.78}Cu_{0.22}$ (NiCu[FM]) layer thickness is taken the same as the NiCu[PM] layer thickness. Note that it is greater than the conventional Co layer thickness since the magnetic moment of NiCu is smaller\cite{Kuru, Nie}.

The samples are fabricated by magnetron sputtering on a sapphire substrate. The base pressure in the chamber does not exceed $10^{-7} Torr$. Before deposition the substrate surface is subjected to mild ionic cleaning. During deposition the substrate is rotated at the rate of $70 rpm$ to obtain a uniform film thickness. The deposition is performed at room temperature in the argon medium with the working pressure of $1.5 - 4 \times 10^{-3} Torr$. The NiCu alloys are made by co-sputtering from two sources. We control the alloy composition by varying the power of magnetrons. The sputtering speed dependence on the magnetron power is pre-measured for Ni and Cu sources separately by atomic force microscope and optical profilometer. The atomic alloy composition is calculated using this data and the tabular values of the Ni and Cu densities and atomic mass. We vary this composition in order to find an alloy that is paramagnetic at room temperature and an alloy with close composition that is ferromagnetic at room temperature. As a result, we obtain $Ni_{0.78}Cu_{0.22}$ as the NiCu[FM] layer and $Ni_{0.68}Cu_{0.32}$ as the NiCu[PM] layer.

In order to check that our chosen alloy compositions correspond to the ferromagnetic and paramagnetic states, respectively, we perform spin-polarized first-principles calculations performed for the Co/$Ni_{0.68}Cu_{0.32}$ and $Ni_{0.78}Cu_{0.22}$/$Ni_{0.68}Cu_{0.32}$ interfaces within density functional theory using periodic supercell models, a plane-wave basis, and the projector augmented-wave formalism (detailed in the supplementary material, Sec. S1, Fig. S1). The calculations reveal a clear difference in the magnetic behavior of the two NiCu alloys: the $Ni_{0.68}Cu_{0.32}$ region relaxes to a state with vanishingly small local magnetic moments and no stable ferromagnetic solution, consistent with its paramagnetic character, while $Ni_{0.78}Cu_{0.22}$ retains finite spin polarization and robust local moments, confirming its ferromagnetic nature. Notably, this contrast is reproduced within conventional spin-polarized DFT, without introducing any additional correlation corrections or empirical parameters.

THz signal $\Delta S(t)$ is measured by a standard terahertz time domain spectroscopy technique described in detail elsewhere \cite{Buryakov2022, Buryakov2023}. The pump beam of a Ti:Sapphire laser (wavelength: 800~nm, pulse duration: 50~fs, pulse repetition rate: 3~kHz) is focused onto the top or bottom side of the sample, depending on whether the measurements from film side or substrate side are performed. The emitted THz pulse is collected by a parabolic mirror and steered onto a 1~mm ZnTe detector crystal. A synchronized probe beam provides electro-optic sampling of the THz waveform. The optical pump fluence (energy density) is varied in the range between $0.1$ and $2.5 \frac{mJ}{cm^2}$, and we control its polarization using a half-wave plate and Glan-Taylor prism. The beam spot size at the sample is equal to 3~mm. The external magnetic field of up to 4~kOe created by an electromagnet is applied in the sample plane along the $Y$-coordinate of the laboratory reference frame. A wire-grid polarizer (WGP) allows polarization-resolved detection of the THz signal. We mainly detect the $X$-polarization that is perpendicular to the applied magnetic field (unless otherwise specified). Additionally, we heat the samples by hot air of different temperature. The sample temperature is controlled by a thermocouple placed right at the sample. This allows us to perform measurements in the temperature range from $20^\circ C$ to $80^\circ C$ with a step of $10 \pm 2 ^\circ C$.

\section{\label{Exp}Experimental results and discussion}
First, we perform the  terahertz time-domain spectroscopy (THz-TDS) of our NiCu THz emitters and compare four different combinations, including NiCu[FM](6~nm) or Co(3~nm) as a ferromagnetic layer and NiCu[PM](6~nm) or Pt(3~nm) as a nonmagnetic layer. We discuss the magnetic properties of the samples. Next, we make THz-TDS for the two NiCu[FM] samples at a $10 s$ time interval with a $3 kHz$ optical pump pulses repetition rate. An independent measurement of THz emission from these samples heated to a $20-80^\circ C$ temperature are made. We compare these two measurements to each other and estimate the temperature of the samples under optical pump of up to $2.5 \frac{mJ}{cm^2}$ fluence.

\subsection{THz signal comparison}
\begin{figure}[tb]
\centering{\includegraphics[width=\hsize]{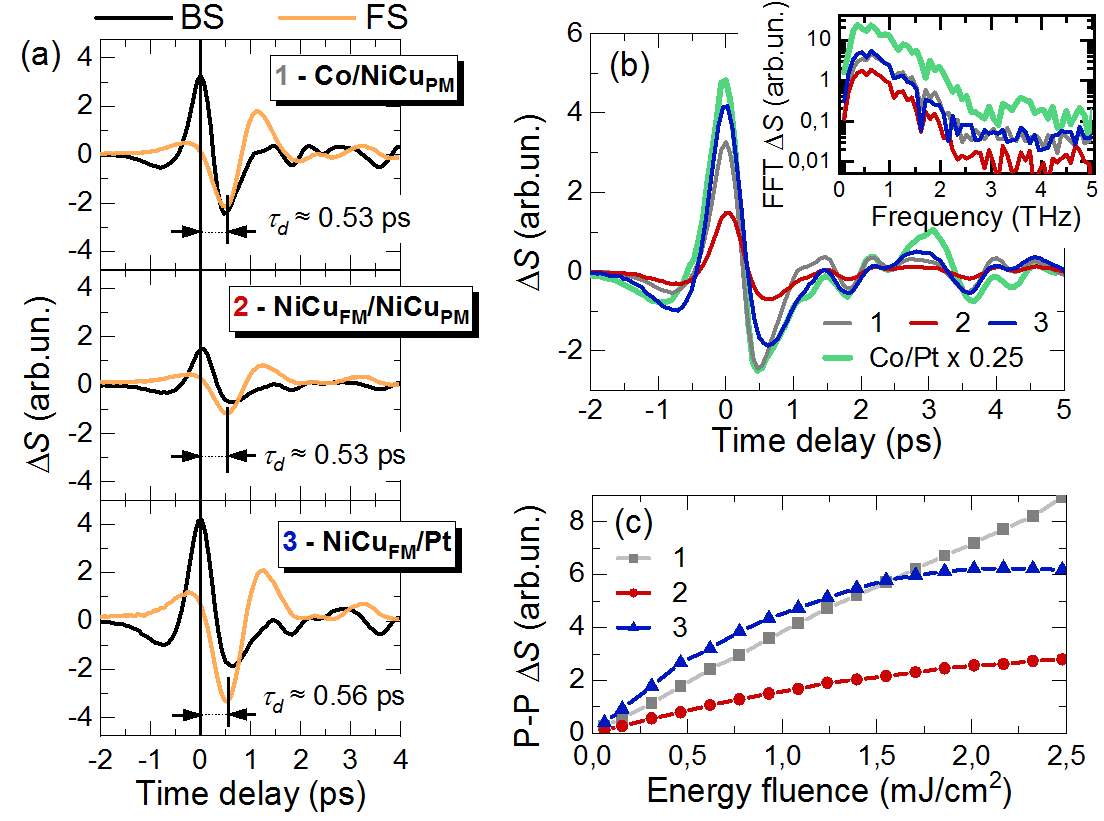}}
\caption{\label{fig1} THz signal from NiCu-based samples under a (nearly) saturating magnetic field of 2~kOe. (a) Time-domain signal for Co/NiCu[PM] (sample 1), NiCu[FM]/NiCu[PM] (sample 2), and NiCu[FM]/Pt (sample 3) irradiated by optical pump from the film side (FS) or substrate side (BS). (b) Comparison of time-domain signal and spectrum (inset) for three samples and standard Co/Pt (divided by 4). (c) Peak-to-peak THz signal versus the fluence of optical pump pulses.}
\end{figure}

The time dependence of the THz signal generated by the structures under investigation is shown in Figure~\ref{fig1}~(a) and (b). The optical pump fluence is set to $1.55 \frac{mJ}{cm^2}$.
One can see that for all our samples the peaks are inverted (signal phase is changed by $180^\circ$) which evidently proves that the THz signal is generated due to the spin current from ferromagnetic layer to nonmagnetic one. This spin current is then converted into an electric one due to spin-orbit coupling, most likely the inverse spin Hall effect. The time delay of $0.53-0.56ps$ between the first THz peaks (shown in Figure~\ref{fig1}~(a)) is due to the difference in optical path and phase velocity for both the optical and terahertz waves.

The comparison of samples 1-3 and the reference Co/Pt sample is shown in Figure~\ref{fig1}~(b). Obviously, the Co/Pt sample shows the greatest THz field magnitude (its amplitude is reduced by a factor of $4$ in the figure for clarity). The NiCu[FM]/Pt sample has a $4.5$ times smaller peak signal than the reference Co/Pt sample which is not surprising taking into account that even clean Ni has the saturation magnetization $M_s$ approximately $0.6$ part of the Co one. We may roughly estimate the saturation magnetization of NiCu[FM] from our experimental data. Supposing a linear dependence of the spin current on $M_s$ \cite{Kampfrath2013, Bull2021} and that for Co $M_{s\,Co} = 1420 G$, we get $M_{s\,NiCu[FM]} \approx 316 G$ at room temperature. This is reasonable for $22 \%$ of copper in the alloy taking into account that $Ni_xCu_{1-x}$ is nonmagnetic at room temperature for $x=0.32$. The fact that the saturation magnetization is smaller for NiCu alloys than that for Co is also known from literature \cite{Kuru, Nie}. Note that our estimation is very rough (by the order of value) since it does not take into account e.g. the electron spin polarization at the Fermi level in the ferromagnet and the boundary impedance.

The Co/NiCu[PM] structure is $6.5$ times worse than the reference Co/Pt structure which leads us to a conclusion that the effective spin Hall angle $\theta_{SH\,NiCu[PM]} \approx \theta_{SH\,Pt} / 6.5$ which is quite good \cite{Keller} as e.g. for Ta it is approximately ten times smaller \cite{Mikhailovsky}. However this strongly depends on the boundary properties as we see from the result for the NiCu[FM]/NiCu[PM] structure. Indeed, from previous estimations one should expect a $4.5 \cdot 6.5 \approx 29$ times decrease in THz peak signal due to the decrease of both the spin current from an FM source and the spin Hall angle of an NM layer. However we observe only a $13$ times decrease in the THz signal. This is explained by the fact that a good boundary match provides approximately $2.3$ times greater spin current injected into the NiCu[PM] spin converter from the NiCu[FM] ferromagnet. We should also note that phase match for samples 2 and 3 corresponds to the same sign of spin Hall angle for Pt and NiCu[PM]. The frequency spectra (shown in the inset in Figure~\ref{fig1}~(b)) are identical for all samples and are limited from above by the value of $2.7 THz$ which corresponds to the sensitivity range of a ZnTe detector.

Figure~\ref{fig1}~(c) shows the dependence of peak-to-peak (i.e. the difference between the positive peak and the following negative peak, contrary to the first peak analyzed in previous paragraph) THz signal generated by samples 1-3 on the optical pump fluence. One can see that the Co/NiCu[PM] sample shows a linear growth in the $0.1-2.5 \frac{mJ}{cm^2}$ fluence range which governs the fact that laser-induced demagnetization is negligible for this sample (the Curie temperature of Co is much greater than the temperature at which the sample is heated by pump). However both NiCu[FM]/NiCu[PM] and NiCu[FM]/Pt samples demonstrate a trend to saturation. This is explained by a concurrence between an increase of the number of spin-polarized carriers excited by pump and decrease of the saturation magnetization of NiCu[FM] due to sample heating. Thus the Curie temperature of our NiCu[FM] is much smaller than that of Co.

\begin{figure}[tb]
\centering{\includegraphics[width=\hsize]{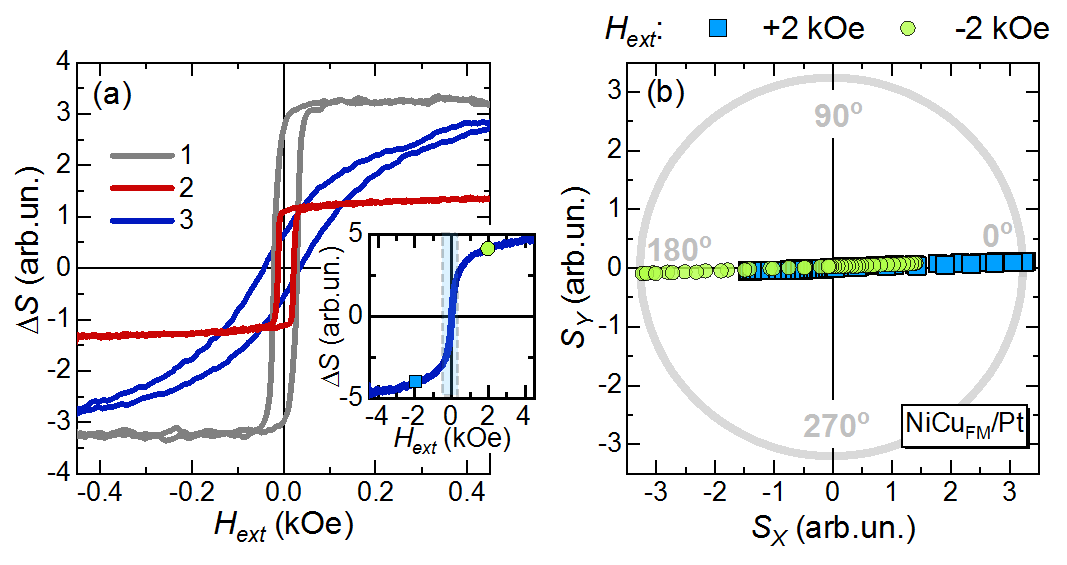}}
\caption{\label{fig2} (a) THz  hysteresis loops obtained via terahertz spintronic magnetometry for Co/NiCu[PM] (sample 1, gray line), NiCu[FM]/NiCu[PM] (sample 2, red line), and NiCu[FM]/Pt (sample 3, blue line) irradiated by optical pump from the film side (FS). The inset shows a hysteresis loop for sample 3 measured in a wider field range of $\pm 4 kOe$ (the field range corresponding to the main picture is shown by blue bar). (b) The $S_X$-$S_Y$ Lissajous curves obtained for sample 3 by independent measurements of two orthogonal components $\Delta S_X$ and $\Delta S_Y$ under the applied magnetic field $H_{ext} = +2 kOe$ (blue rectangles) and $H_{ext} = -2 kOe$ (green rounds).  The gray circle is a guide for eyes.}
\end{figure}

The terahertz hysteresis loops for Co/NiCu[PM] (sample 1), NiCu[FM]/NiCu[PM] (sample 2), and NiCu[FM]/Pt (sample 3) obtained at a fixed delay between pump and probe pulses (corresponding to the peak THz signal) are shown in Figure~\ref{fig2}~(a). 
The external magnetic field $H_{ext}$ is directed along the $Y$-axis in the sample plane. As one can see, a change in the polarity of $H_{ext}$ leads to a change in the sign of the THz signal for all three structures due to the reversal of the direction of magnetization (and the sign of the spin current). Almost rectangular hysteresis loops with coercive fields of $H_c \approx 25 Oe$ and $H_c \approx 17 Oe$, respectively, are observed for Co/NiCu[PM] (sample 1) and NiCu[FM]/NiCu[PM] (sample 2). However for NiCu[FM]/Pt (sample 3), the shape of the THz hysteresis loop (weak hysteresis and small signal in the absence of magnetic field) and large saturation field of $4 kOe$ (inset in Figure~\ref{fig2}~(a)) indicate the formation of a perpendicular magnetic anisotropy due to a strong perpendicular surface anisotropy at the boundary of NiCu[FM] induced by Pt. This is confirmed by MOKE measurements (see supplementary material, Sec.S2).

Although the field $H_{ext} = \pm 2 kOe$ is far from saturating for this structure it allows to achieve approximately $80 \%$ of the maximum THz signal because the magnetization is directed mainly along the sample plane. An additional analysis of the polarization of THz radiation is performed for these field values of field for the NiCu[FM]/Pt sample (sample 3; marked with colored dots in the inset of Figure~\ref{fig2}~(a)). For the two orthogonal components of the THz field $\Delta S_X$ and $\Delta S_Y$ (the X and Y axes are introduced in Section~\ref{Setup}), time dependencies were measured. In order to isolate each component, the axis of the ZnTe detector crystal [-110] was alternately oriented parallel to either the $X-$ or$Y-$axis. The Lissajous curves constructed from these data, shown in Figure~\ref{fig2}~(b), make it possible to characterize the polarization state of the THz wave. The ratio of the amplitudes of the components is $\frac{\Delta S_x}{\Delta S_Y} \approx 30$, and their phase difference is zero. This indicates that the generated THz radiation is almost completely linearly polarized along the X-axis (deviation is less than $1-2^\circ$), and the contribution of the orthogonal component is negligible. When an external magnetic field reverses (from $+2 kOe$ to $-2 kOe$), the direction of the magnetization vector and the associated effective charge current changes to the opposite, which leads to a $180^\circ$ reversal of the polarization of the THz wave without noticeable ellipticity. Thus, the NiCu[FM]/Pt sample under investigation may act as a good source of linearly polarized THz radiation which can be precisely controlled by an external magnetic field.

\subsection{Temperature measurements}
\begin{figure}[t]
\centering{\includegraphics[width=\hsize]{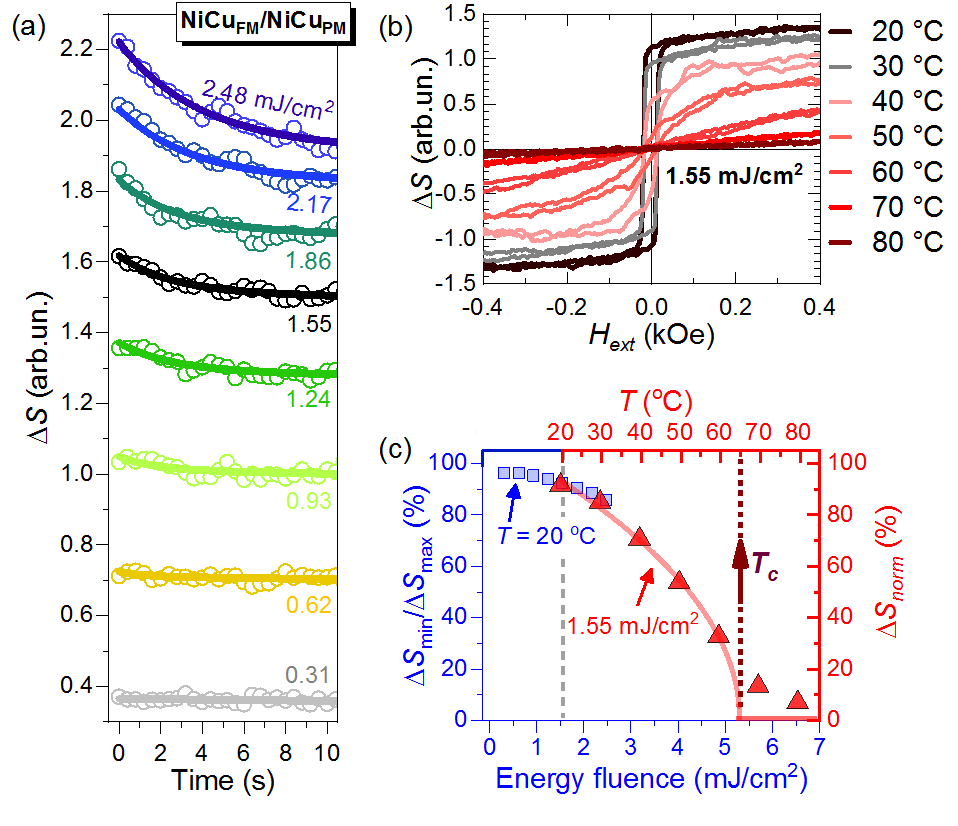}}
\caption{\label{fig3} The influence of laser-induced or thermal heat on the THz emission properties of NiCu[FM]/NiCu[PM] (sample 2). (a) Time dependence of the peak THz amplitude $\Delta S$ for the optical pump fluence in the range of $0.31-2.48 \frac{mJ}{cm^2}$ measured at room temperature ($20^\circ C$) at a fixed pump-probe delay in a $10s-30s$ regime of measurement time and pause between different fluence (in ascending order). Circles correspond to the experimental data, solid lines are two-exponent approximations. (b) Temperature dependence of THz peak-to-peak hysteresis loops measured for a heated sample at a fixed pump fluence of $1.55 \frac{mJ}{cm^2}$. (c) The combined results for laser-induced heating (THz amplitude decrease in percent, blue rectangles, left axis) and thermal heating (THz signal in saturation normalized to the maximum value, red triangles, right axis) of the sample. The red solid line is a theoretical approximation for ferromagnetic-to-paramagnetic transition; the Curie temperature is shown by red arrow. The dashed gray line corresponds to the same conditions for two experiments.}
\end{figure}

In what follows, we show the results of temperature-dependent THz properties of the two samples, NiCu[FM]/NiCu[PM] (sample 2) and NiCu[FM]/Pt (sample 3), heated either by optical pump pulses or by hot air.
Figure~\ref{fig3}~(a) shows the dependence of the peak amplitude of the THz signal on the irradiation time for the NiCu[FM]/NiCu[PM] structure with a sequential change in optical energy fluence in the range from $0.31$ to $2.48 \frac{mJ}{cm^2}$. The experiment is conducted at room temperature ($20^\circ C$). Before the measurement with each fluence, the pump beam is blocked for $30s$, then the beam is opened in the time interval from 0 to 10 seconds, and the amplitude evolution is recorded with a fixed pump-probe delay time. The repetition rate of the pulses is $3 kHz$. Two complete measurement cycles are made in order to check the reproducibility of the results.

We see from Figure~\ref{fig3}~(a) that two effects are observed as the pump fluence increases. First, there is an increase in the initial peak amplitude of the THz signal, followed by saturation, which is consistent with the dependence in Figure~\ref{fig1}~(c). Second, a gradual decrease in the amplitude is seen during each 10-second interval until reaching a quasi-stationary value corresponding to thermal equilibrium at each optical fluence. The solid curves in Figure~\ref{fig3}~(a) are the result of a phenomenological two-exponential approximation which clearly reflects the presence of both fast and slow relaxation components. This behavior demonstrates the competition of two main processes: an increase in generation of the spin-polarized carriers due to the absorption of laser pump and a decrease of the NiCu[FM] magnetization  as a result of laser-induced demagnetization due to heating.

Figure 3(b) shows the THz hysteresis loops for the same  sample 2 measured at different temperatures and a constant optical energy density of $1.55 \frac{mJ}{cm^2}$. One can see that the THz signal decreases as the temperature grows and is close to zero at $80^\circ C$. The temperature dependence of the THz amplitude in saturation normalized to a maximum achieved at room temperature $\Delta S_{norm}(T)$ for the smallest fluence (the procedure of normalization is described below) is plotted in Figure~\ref{fig3}(c) (red triangles). The temperature dependence of the terahertz amplitude is mainly determined by a linear dependence in $M_s\left(T\right)$. Approximation $\Delta S_{norm}(T) = C \left(1 – \frac{T}{T_C}\right)^{0.5}$ with the use of standard expression for magnetization near the Curie temperature gives an estimate of $T_C \approx 65 ^\circ C$ for the ferromagnetic NiCu layer (we estimate the error as $\pm 2 ^\circ C$).

The blue squares in Figure~\ref{fig3}~(c) correspond to a relative decrease in the amplitude of the THz signal of $100\% \cdot \frac{\Delta S_{min}}{\Delta S_{max}}$ (at a $10 s$ time interval) extracted from the data shown in Figure~\ref{fig3}~(a). Here $\Delta S_{max}$ is the amplitude of the THz signal at the time the laser pump is turned on and $\Delta S_{min}$ is the amplitude after thermal equilibrium is achieved (approximately $10 s$). We observe approximately $15\%$ decrease $100\% \cdot \left(1 - \frac{\Delta S_{min}}{\Delta S_{max}}\right)$ at the maximum optical pump fluence. The dashed gray line connecting the marks in Figure~\ref{fig3}~(c) at $T = 20^\circ C$ and an energy density of $1.55 \frac{mJ}{cm^2}$ allows us to match the results of two independent experiments with homogeneous thermal heating and laser-induced heating under identical conditions (this point is then taken as a reference for obtaining $\Delta S_{norm}(T)$, and thus we achieve a signal normalized to the signal obtained for the smallest value of fluence under which the laser-induced heating is negligible). The comparison shows that the optical pump with the fluence of $2.5 \frac{mJ}{cm^2}$ heats the sample by about $10^\circ C$. This estimate confirms that prolonged exposure to high-fluence laser pulses leads to significant laser-induced demagnetization of the NiCu[FM] layer.

\begin{figure}[t]
\centering{\includegraphics[width=\hsize]{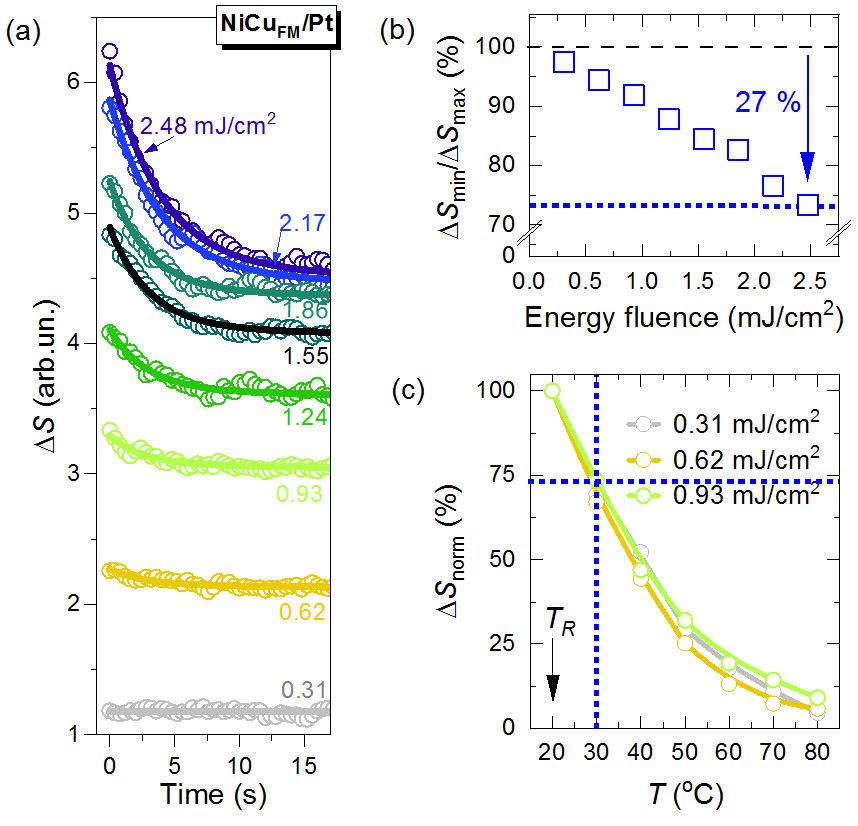}}
\caption{\label{fig4} The influence of laser-induced or thermal heat on the THz emission properties of NiCu[FM]/Pt (sample 3). (a) Time dependence of the peak THz amplitude $\Delta S$ for the optical pump fluence in the range of $0.31-2.48 \frac{mJ}{cm^2}$ measured at room temperature ($20^\circ C$) at a fixed pump-probe delay in a $15s-30s$ regime of measurement time and pause between different fluence (in ascending order). Circles correspond to the experimental data, solid lines are two-exponent approximations. (b) The THz amplitude decrease in percent due to laser-induced heating of the sample for $15s$. The dashed line and arrow show maximum obtained THz signal decrease of $27 \%$. (c) The THz signal decrease due to thermal heating of the sample under an applied saturating magnetic field of $4 kOe$ normalized to the maximum value obtained at room temperature measured for three different optical pump fluences ($0.31, 0.62, 0.93 \frac{mJ}{cm^2}$). The room temperature is shown by arrow. The dashed line shows the $\Delta S$ value corresponding to the maximum laser-induced heating and temperature corresponding to it.}
\end{figure}

Similar measurements were performed for the NiCu[FM]/Pt structure (sample 3); however the laser exposure time was increased to $15 s$ in order to reach saturation. The results are shown in Figure~\ref{fig4}. In contrast to sample 2, an increase in optical energy density for NiCu[FM]/Pt is accompanied by a sharper saturation in the amplitude of the THz signal, which is consistent with the data in Figure~\ref{fig1}~(c). The time evolution of $\Delta S(t)$ at the maximum energy density ($2.48 \frac{mJ}{cm^2}$) demonstrates significant laser‑induced demagnetization: the decrease in the ratio $100\% \cdot \left(1 - \frac{\Delta S_{min}}{\Delta S_{max}}\right)$ is $27\%$ (see Figure~\ref{fig4}~(b)) which is almost two times greater than that for the NiCu[FM]/NiCu[PM] structure.

The temperature dependencies of the normalized amplitude $\Delta S_{norm}(T)$ obtained at three different optical energy densities (0.31, 0.62, and 0.93 $\frac{mJ}{cm^2}$) are shown in Figure~\ref{fig4}~(c). They practically coincide after normalization to the maximum which indicates a single thermally induced mechanism of magnetization drop (through the ferromagnet-paramagnet phase transition). As in the case of sample 2, a comparison of the dependencies in Figure~\ref{fig4}~(b) and (c) allows us to estimate that the maximum energy density ($2.48 \frac{mJ}{cm^2}$) is equivalent to sample heating by approximately $10^\circ C$. Here we observe a $27\%$ decrease $100\% \cdot \left(1 - \frac{\Delta S_{min}}{\Delta S_{max}}\right)$ in the terahertz signal at the maximum optical pump fluence. It seems from our measurements that the Curie temperature of the NiCu[FM]/Pt structure is slightly less than that of the NiCu[FM]/NiCu[PM] structure: it may be estimated as $60 \pm 2 ^\circ C$. However the difference of the Curie temperature of NiCu[FM] roughly lies within the error interval.

Thus we show that the heterostructures based on NiCu[FM] allow to realize a thermoinduced control of its magnetization which can be used to manipulate by the THz emission rate by either laser-induced or thermal heating of the sample. Besides, the amplitude of linearly polarized THz radiation may be effectively controlled by an external magnetic field for the NiCu[FM]/Pt structure.

\section{\label{Concl} Conclusion}
We investigate the spintronic THz emitters in which both ferromagnetic and nonmagnetic layers are NiCu alloys with different composition, one being in the ferromagnetic state and the other in paramagnetic state. We show that although the saturation magnetization of NiCu[FM] is much smaller than that of Co the peak THz signal generated by it is only $2$ times smaller if NiCu[PM] is chosen as a spin-to-charge converter (for the pump fluence of $1.55 \frac{mJ}{cm^2}$). On the other hand, the NiCu[PM] alloy is only $2.9$ times worse than Pt as a spin-to-charge converter if the spin source is NiCu[FM]. The etalon Co/Pt structure is $4.5$ times better than the NiCu[FM]/Pt structure. A simple comparison of the mentioned numbers leads to the conclusion that the effectiveness of our THz emitters cannot be described as the simple product of the efficiencies of the spin source and spin-to-charge converter of spin source and spin-to-charge converter. Here we show that the boundary match in the structure in which NiCu with different alloy composition acts as both FM and NM layer leads to the $2.3$ times growth of THz signal.

Another important result is related to the relatively low Curie temperature of the NiCu[FM] layer, estimated to be about $65^\circ C$. This makes NiCu-based emitters attractive for thermal control of the THz response. By comparing laser-induced and externally induced heating, we estimate that the maximum optical pump fluence used in our experiments corresponds to an effective temperature increase of about $10^\circ C$, which is sufficient to noticeably reduce the THz emission amplitude. In addition, the NiCu[FM]/Pt structure emits nearly linearly polarized THz radiation for which the amplitude and polarity can be controlled by an external magnetic field.

These results demonstrate that NiCu-based spintronic THz emitters are a promising platform for THz generation in applications where moderate output amplitude can be combined with thermal tunability and interface-engineered functionality.

\begin{acknowledgments}
This work was supported by the Ministry of Science and Higher Education (Contract No. FSFZ-2025-0002). The experimental measurements were carried out with support from RSF Grant No. 25-79-00247.
\end{acknowledgments}

\nocite{*}

\bibliography{apssamp}

\end{document}